\documentclass[10pt,conference]{IEEEtran}
\usepackage{cite}
\usepackage{amsmath}
\usepackage{amssymb}
\usepackage{algorithmicx}
\usepackage{algpseudocode}
\usepackage{paralist}
\usepackage{xcolor}
\usepackage{xspace}
\usepackage{listings}
\usepackage{hyperref}
\usepackage{cleveref}
\usepackage{latexsym}
\usepackage{colortbl}
\usepackage{graphicx}
\usepackage{comment}
\usepackage{booktabs}
\usepackage{multirow}

\usepackage{tikz}
\usetikzlibrary{positioning}

\newcommand\SI[1]{{\color{purple}{SI: #1}}}
\newcommand\ES[1]{{\color{blue}{ES: #1}}}

\newcommand\eg{\textit{e.g.}\@\xspace}
\newcommand\ie{\textit{i.e.}\@\xspace}
\newcommand\vs{\textit{vs.}\@\xspace}

\newcommand\funcname[1]{\textsf{\fontsize{9pt}{8pt}\selectfont#1}}
\newcommand\smallfuncname[1]{\textsf{\fontsize{7.5pt}{7pt}\selectfont#1}}

\newcommand\tfilter{\funcname{filter}}
\newcommand\tmax{\funcname{max}}
\newcommand\tmin{\funcname{min}}
\newcommand\ttrue{\funcname{true}}
\newcommand\tfalse{\funcname{false}}

\newcommand\tsmallmax{\smallfuncname{max}}
\newcommand\tsmallmin{\smallfuncname{min}}

\newcommand\tif{\mathrm{if}}
\newcommand\tthen{\mathrm{then}}
\newcommand\telse{\mathrm{else}}

\newcommand\cblue{{\color{blue}blue}\xspace}
\newcommand\cred{{\color{red}red}\xspace}

\newcommand\rwto{\overset{.\,}{\rightarrow}}

\newcommand\congblue{{~\color{blue}\cong_{\mathrm{b}}\,\,}}
\newcommand\congred{{~\color{red}\cong_{\mathrm{r}}\,\,}}

\newcommand\myparagraph[1]{
  \smallskip\noindent\textbf{#1}~}

\lstdefinestyle{codestyle}{
    keywordstyle={\bfseries}, 
    commentstyle={\color{codegreen}},
    stringstyle=\color{codepurple},
    basicstyle=\ttfamily\footnotesize,
    numbers=left, 
    language=Python 
} 
\begin{document}
\title{Colored E-Graph: Equality Reasoning with Conditions}
\author{
    \IEEEauthorblockN{Eytan Singher}
    \IEEEauthorblockA{
    Technion\\
    eytan.s@cs.technion.ac.il
    }
    \and
    \IEEEauthorblockN{Shachar Itzhaky}
    \IEEEauthorblockA{
    Technion\\
    shachari@cs.technion.ac.il
    }
}
\maketitle
\thispagestyle{plain}
\pagestyle{plain}
\begin{abstract}

E-graphs are a prominent data structure that has been increasing in popularity in recent years due to their expanding range of applications in various formal reasoning tasks. 
Often, they are used for equality saturation, a process of deriving consequences through repeatedly applying universally quantified equality formulas via term rewriting.
They handle equality reasoning over a large spaces of terms, but are severely limited in their handling of case splitting and other types of logical cuts, especially when compared to other reasoning techniques such as sequent calculi and resolution.
The main difficulty is when equality reasoning requires multiple inconsistent assumptions to reach a single conclusion.
Ad-hoc solutions, such as duplicating the e-graph for each assumption, are available, but they are notably resource-intensive.


We introduce a key observation is that each duplicate e-graph (with an added assumption) corresponds to coarsened congruence relation.
Based on that, we present an extension to e-graphs, called \emph{Colored E-Graphs}, as a way to represent all of the coarsened congruence relations in a single structure.
A colored e-graph is a memory-efficient equivalent of multiple copies of an e-graph, with a much lower overhead.
This is attained by sharing as much as possible between different cases, while carefully tracking which conclusion is true under which assumption.
Support for multiple relations can be thought of as adding multiple ``color-coded'' layers on top of the original e-graph structure, leading to a large degree of sharing.

In our implementation, we introduce optimizations to rebuilding and e-matching.
We run experiments and demonstrate that our colored e-graphs can support hundreds of assumptions and millions of terms with space requirements that are an order of magnitude lower, and with similar time requirements.
\end{abstract}

\section{Introduction}
\label{intro}

E-graphs are a versatile data structure that is used for various tasks of automated reasoning, including theorem proving and synthesis.
They are especially effective in reasoning about equality.
E-graphs have been popularized in compiler optimizations thanks to their ability to support efficient \emph{rewrites} over a large set of terms, while keeping a compact representation of all possible rewrite outcomes.
This mechanism is known as \emph{equality saturation}.
It provides a powerful engine that allows a reasoner to generate all equality consequences of a set of known, universally quantified equalities.
Possible uses include selecting the best version of an expression according to some desired metric, such as run-time efficiency~\cite{POPL2009:Tate}, size~\cite{DBLP:conf/fmcad/FlattCWTP22,DBLP:conf/fmcad/NotzliBNPRBT22}, or precision~\cite{herbie}
(when used as a compilation phase)
and a generalized form of unification, called e-unification, for application of inference steps (when used for proof search).

In this work we focus on \emph{exploratory reasoning} tasks such as theory exploration \cite{thesy}, rewrite rule inference \cite{DBLP:journals/pacmpl/NandiWZWSASGT21}, and proof search \cite{CAV2013:Kovacs,DBLP:conf/aplas/BrotherstonGP12,cycleq}.
Exploratory reasoning involves some intermediate (or even final) conclusions or goals that are not known a priori, and need to be discovered.
For example, in the course of theory exploration, we traverse the domain of list expressions to find all valid equalities.
One of the candidate equalities encountered may be:
\[\tfilter~q~(\tfilter~p~[x,y]) =
  \tfilter~p~(\tfilter~q~[x,y])\]

Where the definition of $\tfilter$ is
\[\tfilter~p~(x::xs) = 
\tif~p\,x~\tthen~x::\tfilter~p~xs 
~\telse~\tfilter~p~xs\]

We wish that the automatic reasoning technique simplify both terms to a common form, in order to prove that they are equal.
A term like $\tfilter~p~[x,y]$ is rewritten using the definition to 
$\tif~p\,x~\tthen~x::\tfilter~p~[y]~\telse~\tfilter~p~[y]$.
In this example we assume an equality saturation engine, based on an e-graph, that can simplify terms of the form $\tif~\ttrue...$ and $\tif~\tfalse...$, but cannot directly rewrite $\tif~p\,x...$ because $p\,x$ does not match either $\ttrue$ or $\tfalse$;
therefore, this term requires special treatment.

When encountering a term like $\tif~p\,x$, previous work based on rewriting, specifically equality saturation, either leaves it as opaque, or performs \emph{case splitting} by forking the entire state (in this case, the e-graph).
Case splitting, a common reasoning step, is needed to introduce alternative assumptions $p\,x$ and $\lnot p\,x$ and consider each one.
With the forking case splitting method we will have one copy in which 
$\tfilter~p~[x,y]$ rewrites to $x::\tfilter~p~[y]$
and one copy in which 
it rewrites to $\tfilter~p~[y]$.
The two copies will have a shared term
$\tfilter~p~[y]$, but this cannot be taken advantage of, because the copies are disjoint;
therefore any rewrite that applies to the term will now have to be carried out twice, doubling both time and memory consumption.
Furthermore, rewriting of this latter term, $\tfilter~p~[y]$, will similarly have to case-split on $p\,y$,
so each copy will have to be cloned again.
This, even though the rewriting of $\tfilter~p~[y]$ depends only on $p\,y$ and is agnostic to which case of $p\,x$ is taken.
Further rewriting of the terms in the equality above exacerbates problem, because $q\,x$, $q\,y$ also occur in conditions, perpetrating additional case-splits, and leading to even more duplication ($\times16$) as well as redundant, repeated work.
In larger examples, the reasoner will quickly exhaust the available memory as a result.

We note that in exploratory tasks it is often necessary to keep all the exploration paths rather than ``finish off'' proof branches iteratively.
The reason is that an incomplete path can be completed later in the exploration with the help of other conclusions.
This makes the memory limitation acute to these scenarios.

Our key observation is that each assumption may lead to additional unions, but may never ``break apart'' classes of terms.
An e-graph naturally represents a congruence relation $\cong$, which is an equality relation over terms (with function applications), which  maintains $x \cong y \vdash f(x) \cong f(y)$.
Any union done on the assumption-less congruence relation, which we call the ``root'' congruence relation, will also be correct on any forked e-graphs' congruence relation with additional assumptions.
This observation is important because such unions in the root can be shared across all other derived e-graphs.

We extend the e-graph data structure into a \emph{Colored E-Graph} to maintain multiple congruence relations at once, where each relation is associated with a color.
A colored e-graph contains multiple congruence relations without having to duplicate the shared terms, by holding a union-find structure for each relation.
For the example above, we will use a colored e-graph with different colors for assumptions on the truth value of the $\tif$ condition, $p\,x$. 
The root $\cong$ will be represented by the color black.
The color \cblue will represent the assumption $p\,x$, which we denote $p\,x \congblue \ttrue$, and the color \cred will represent $\lnot p\,x$ which we denote $p\,x \congred \tfalse$.
Equality saturation can now infer, via rewriting, the consequences $\tfilter~p~[x,y] \congblue x::\tfilter~p~[y]$ and $\tfilter~p~[x,y] \congred \tfilter~p~[y]$.
When creating the new \cblue and \cred, the term $\tfilter~p~[x,y]$ will not be copied; instead, it is shared with the black color.

We also extend the colored e-graph's logic in several key ways, to support e-graph operations for all the colors.
First, we set up a multi-level union-find where the lowest level corresponds to the root congruence.
Higher levels define further unions of e-class representatives from the lower one, forming \emph{colored e-classes}, giving a coarsening of the root relation.
Second, we change how congruence closure is applied to the individual congruence relations while taking advantage of the sharing between each such relation and the root. 
Lastly, we present a technique for e-matching over all the relations at once.

Our contributions:
\begin{enumerate}
    \item The observation that assumptions induce coarsened e-graphs that share much of the original structure. 
    \item Algorithms for colored e-graphs operations.
    \item Optimizations on top of the basic algorithms to significantly improve resource usage.
    \item A colored e-graph implementation\footnote{\url{https://github.com/eytans/egg/tree/features/color_splits}} and an evaluation that shows an improvement factor in memory usage over the existing baseline, while maintaining similar run-time performance.
\end{enumerate}
\section{Background on E-graphs}
\label{background}

This section presents some basic definitions
and notations that will be used throughout the paper.
We assume a term language $L$ where terms are constructed using \emph{function symbols}, each with its designated arity.
We use $f^{(r)}\in\Sigma[L]$ to say that $f$ is
in the \emph{signature} of $L$ and has arity $r$.
A term is then a \emph{tree} whose nodes are labeled by function symbols and a node labeled by $f$ has $r$ children.
(In particular, the leaves of a term have nullary function symbols.)

An e-graph $\mathcal{G}$ serves as a compact data structure representing a set $S\subseteq L$ of terms and a congruence relation ${\cong}\subseteq L\times L$. This congruence relation, in addition to being reflexive, symmetric, and transitive, is also closed under the function symbols of $\Sigma[L]$. That is, for every $f^{r}\in\Sigma[L]$, and given two lists of terms $t_{1..r}\in L$ and $s_{1..r}$, each of length $r$, if $t_i\cong s_i ~ (i=1..r)$, then it follows that $f(t_1,\ldots,t_r)\cong f(s_1,\ldots,s_r)$. This property, known as \emph{congruence closure}, is a key attribute of the data structure. The maintenance of this attribute as an invariant significantly influences the design and implementation of e-graph actions.

The egg library~\cite{egg} revolutionizes the application of e-graphs by explicitly supporting the equality saturation workflow.
It enables the periodic maintenance of congruence closure, via \emph{deferred rebuild}, allowing for the amortization of associated rebuilding costs.
We give a short background on how egg achieves better performance by means of efficient data structure representation.

In egg, the authors present the e-graph as a union-find like data structure, augmented to support operations on expressions.
This implementation is primarily achieved through the utilization of three key structures: a hash-cons table, a union-find structure, and an e-class map.
These structures collectively underpin the functionalities integral to the operation of the e-graph.
%
(1) The \underline{hash-cons} table maps e-nodes---which are similar to AST nodes except that their children point to e-class ids instead of a sub-term.
An important aspect of the hash-cons is that after rebuilding its keys are expected to be \emph{canonical}: whenever e-classes are merged, one of their ids becomes ``the'' representative id of the new class.
If $x$ and $y$ are merged into $y$, then $f(x,z)$ must become $f(y,z)$.
Same is true for the e-class ids associated with all e-nodes.
%
(2) The \underline{union-find} component responsible for keeping track of merged e-classes and maps each e-class id to a single representative for all (transitively) merged e-classes.
This information is later used to canonicalize the keys and values of the hash-cons.
%
(3) The \underline{e-class map} stores the structure of the e-graph.
For each e-class id, the map keeps all the e-nodes that are contained therein (and through which child e-classes can be reached) and all the parent e-classes, i.e., those containing an e-node one of whose children is that e-class.

For example, the term $a\cdot(b + c)$ may be stored in an e-graph using five e-classes $[1]$ though $[5]$ (square brackets will denote an e-class id in our example) and the following hash-cons:
\[a \mapsto [1]; b\mapsto [2]; c\mapsto [3];
  [2] + [3] \mapsto [4]; [1] \cdot [4]\mapsto [5]\]

\noindent and e-class map:
\[
\begin{array}{l}
[1]\mapsto \{a\}, \{[5]\}; ~~~
  [2]\mapsto \{b\},\{[4]\}; ~~~ [3]\mapsto\{c\}, \{[4]\};\\
{}[4]\mapsto \{[2]+[3]\},\{[5]\}; \quad [5]\mapsto \{[1]\cdot[4]\}, \varnothing
\end{array}
\]

This trivial example is boring because each e-class contains exactly one e-node and no classes are merged.
This is about to change once we start mutating the graph via operations.

(1) \underline{Insert}. Adding a term $t$ to the e-graph basically means creating an e-class per AST node of $t$; but this can potentially create many duplicates as some (or all) sub-terms of $t$ may already be present in the graph.
For this reason, insertion is done bottom-up where 
at each level the hash-cons is utilized to look up an existing, compatible e-node, in which case its containing e-class id is reused.
(2) \underline{Union}. Merging two e-classes is accomplished by applying the respective union operation of the union-find and concatenating the node and parent lists in the e-class map.
This, however, temporarily invalidates the invariant of the hash-cons and e-class map that all e-class ids and e-nodes must be canonical.
(3) \underline{Rebuilding (Congruence closure)}. As explained before, a union of $[x]$ into $[y]$ necessitates replacing any e-node $f([x],[z])$ by $f([y],[z])$.
Moreover, if $f([x],[z])\in[w_1], f([y],[z])\in[w_2]$,
then, following this replacement, both $[w_1]$ and $[w_2]$ now contain
$f([y],[z])$, meaning that $[w_1] = [w_2]$ evoking a cascading union of $[w_1], [w_2]$.
These updates accumulate quickly and are very costly to keep up with during rewriting.
A significant contribution by egg is the concept of deferred rebuilding, where rebuilding is performed periodically.
The periodic rebuilding is highly efficient and well-suited for equality saturation; 
however, it does not provide significant gains for tasks that require the congruence invariant to be maintained at all times.
(4) \underline{E-matching}. Perhaps the most interesting and algorithmically involved operation is looking up a \emph{pattern} in the set of terms represented by the e-graph.
A pattern is a term with (zero or more) \emph{holes} represented by metavariables $?v_{1..k}$.
For example, $(?v_1+1)\cdot ?v_2$ is a pattern.
Pattern lookup is important for rewriting in equality saturation.
Since the metavariable holes may be filled in by any term,
existing matching algorithm operate
in a top-down manner, traversing the e-nodes downward via the e-class map.

\smallskip
\noindent\textbf{Rewriting.~}
We expand a bit about equality saturation and rewriting.
In this setting, we assume a background set of symbolic \emph{rewrite rules} (r.r.), each of the form $t\rwto s$,
where $t$ and $s$ are patterns as explained in item (4) above.
Repetitively applying such rewrite rules to a set of terms can be used to generate growing sets of terms that are equivalent, according to rewrite semantics, to ones in the starting set.
These are added to the e-graph and arranged into equivalence classes.
Ideally, the set eventually \emph{saturates}, in which case the e-graph now describes \emph{all} the
terms that are rewrite-equivalent.
We point out that in many situations, the e-graph keeps growing as a result of rewrites and never gets saturated---so the number of successive rewrite iterations, or ``rewrite depth'', has to be bounded.

A \emph{conditional rewrite rule} (c.r.r.)~\cite{jcss/Bergstra} is a natural extension of a r.r. that has the following form:
\[\varphi \Rightarrow t \rwto s\]

Where $\varphi$ is a precondition for rewriting $t$ to $s$. For example, the rules for $max$ are:
\[
\begin{array}{l}
{?x} > {?y} \Rightarrow \tmax({?x}, {?y}) \rwto {?x} \\
{?x} \leq {?y} \Rightarrow \tmax({?x}, {?y}) \rwto {?y}
\end{array}
\]

The semantics of a precondition $\varphi$ is defined such that a term matching the pattern of $\varphi$
must be unified with Boolean $\ttrue$ in order for the rewrite to be applied.

\section{Colored E-Graphs: Overall Design}
\label{chap:colored-egraph}
\label{overview}


\begin{figure}
  \centering
  \begin{tabular}{ccc}
  $\cong$ & $\congblue$ & $\congred$ \\
  \includegraphics[width=2.5cm]{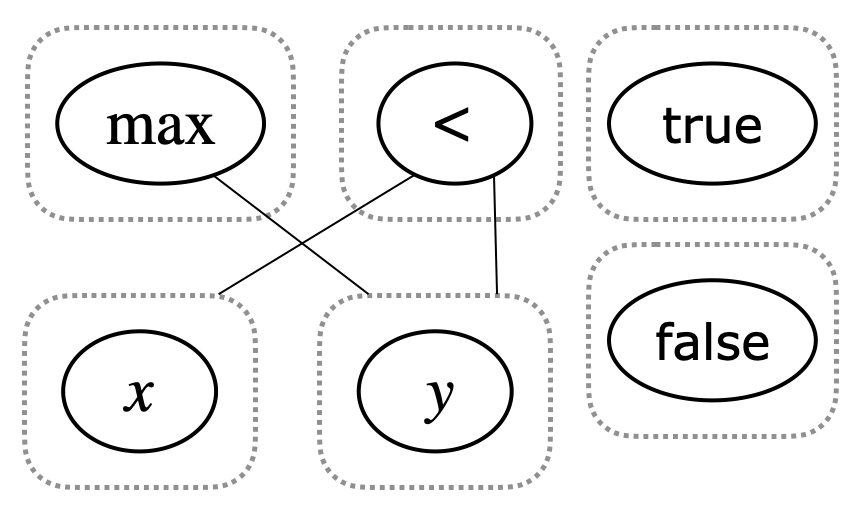}
  &
  \includegraphics[width=2.5cm]{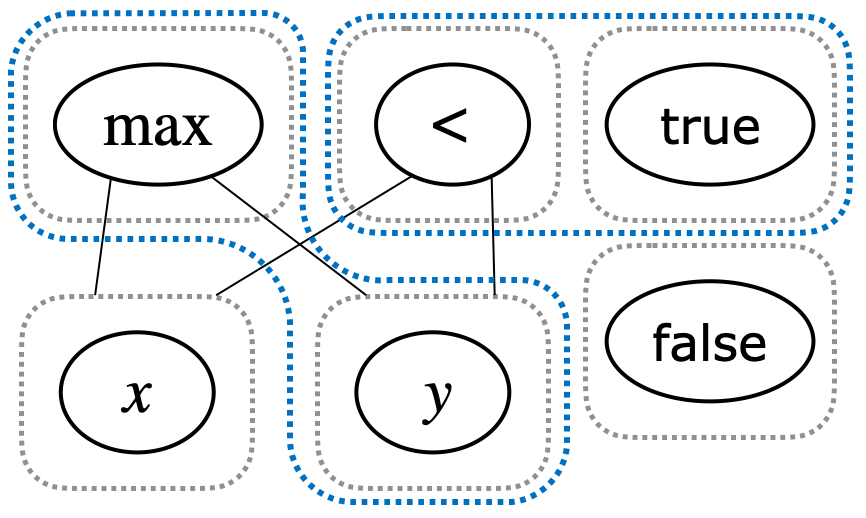}
  &
  \includegraphics[width=2.5cm]{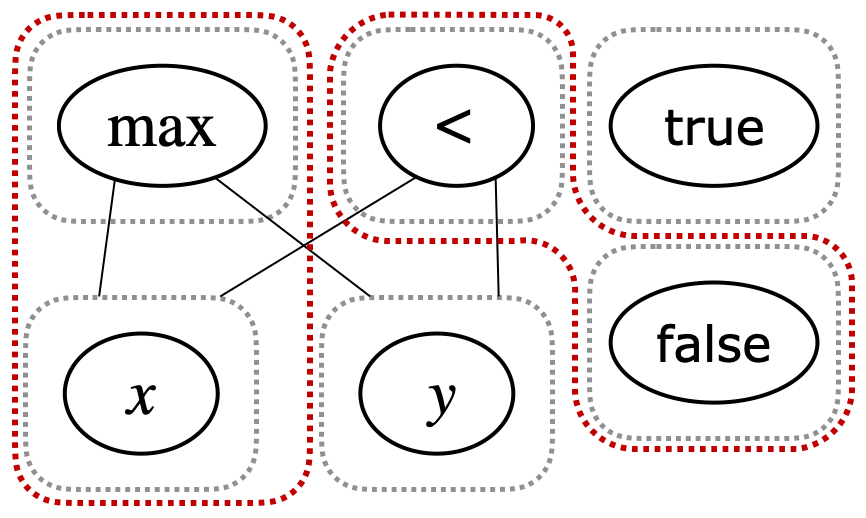}
  \\
  {\small (a)} & {\small (b)} & {\small (c)}
  \end{tabular}

  \medskip
  \begin{tabular}{c}
  \includegraphics[width=3.7cm]{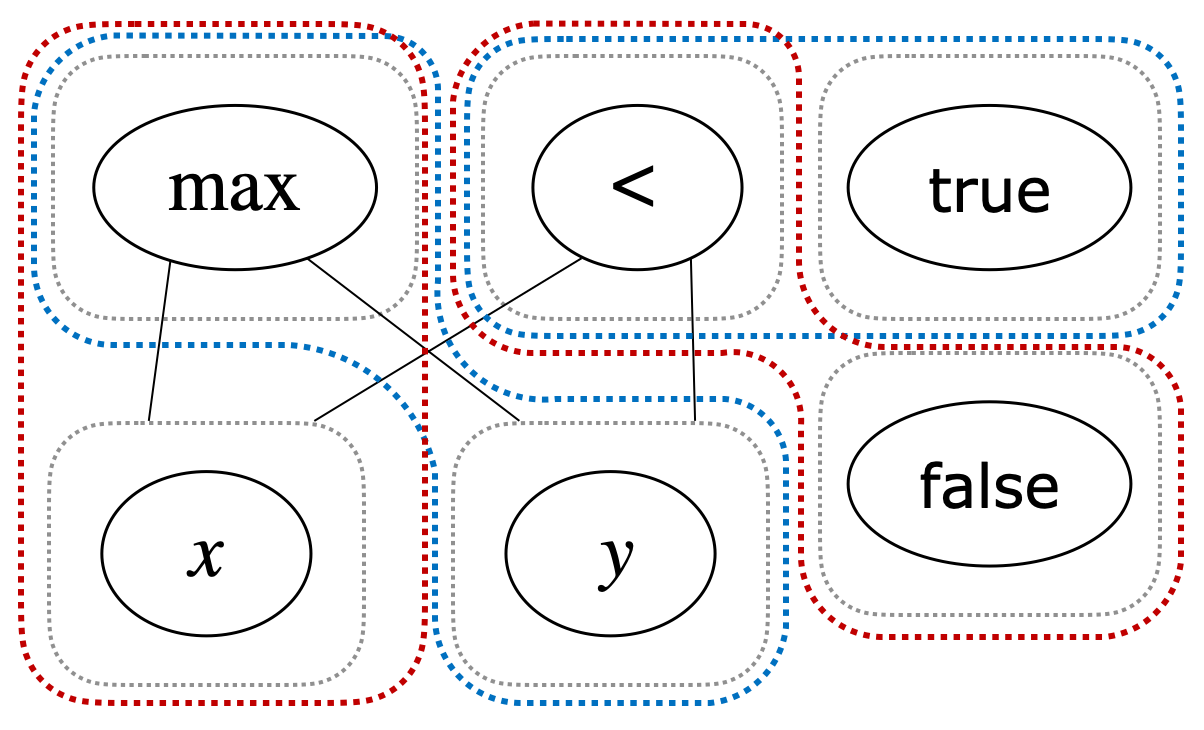} \\
  {\small (d)}
  \end{tabular}
  \caption{Example e-graph (a), with two colored layers; (b) is blue, (c) is red, (d) shows them combined.}
  \label{overview:egraph-max}
\end{figure}

This study aims to extend automatic reasoning through equality saturation, by providing direct support for simultaneous reasoning about multiple cases (or multiple congruence relations).
The main use case for handling multiple relations in parallel arises when attempting to reason about an unknown goal in the presence of conditional expressions. 
Notably, existing systems such as TheSy~\cite{thesy} and Ruler~\cite{DBLP:journals/pacmpl/NandiWZWSASGT21} utilize equality saturation with e-graphs in an exploratory setup. 
In their setup, the primary objective is to discover new rewrite rules that are initially unknown at the start of the execution.
In these exploratory reasoning tasks, we might want to work with additional assumptions. 
A good example is case splitting steps required to handle conditional expressions.
For example, let $t := \tmax(x, y)$, then reasoning about the cases $x > y$ and $x \leq y$ separately is desirable: in the first case $t \cong x$, and in the second $t \cong y$ while without additional assumptions we cannot say either.
The approach in \cite{thesy} uses a prover with a case splitting mechanism which creates a clone e-graph for each such case, $x > y$ and $x \leq y$, but this runs at great expense to run-time and memory.
In particular, the e-graph may consist of a large number of terms that are not related to $x$ and $y$ or their relative values.
Such terms will be duplicated needlessly, and subsequent rewrites that apply to them will be repeated on the duplicates.
In cases where further case splitting is required, e.g. $y > z$, that split must also be done on both clones, leading to a duplication factor of 4.
Thus, the number of clones will grow exponentially with the number of nested case splits.

\begin{figure}
  \centering
  \begin{tabular}{@{}c@{~}c@{~}c@{}}
    $\cong$ & $\congblue$ & $\congred$ \\
    \includegraphics[width=0.15\textwidth]{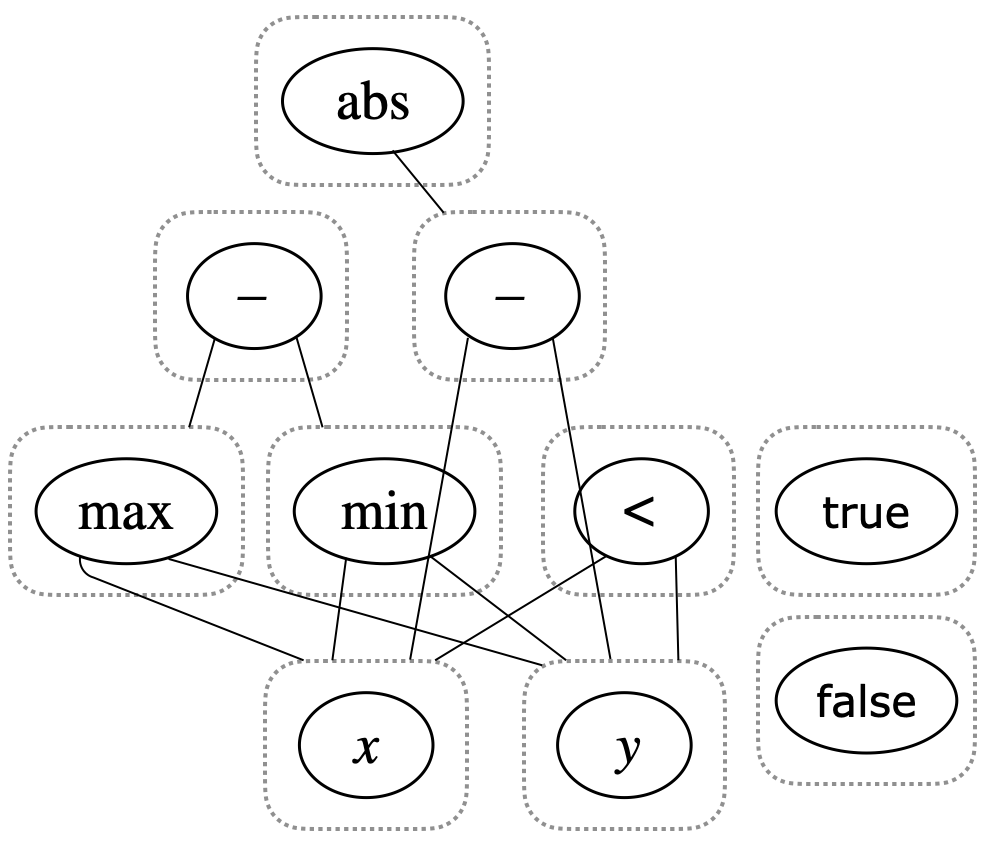} &
    \includegraphics[width=0.15\textwidth]{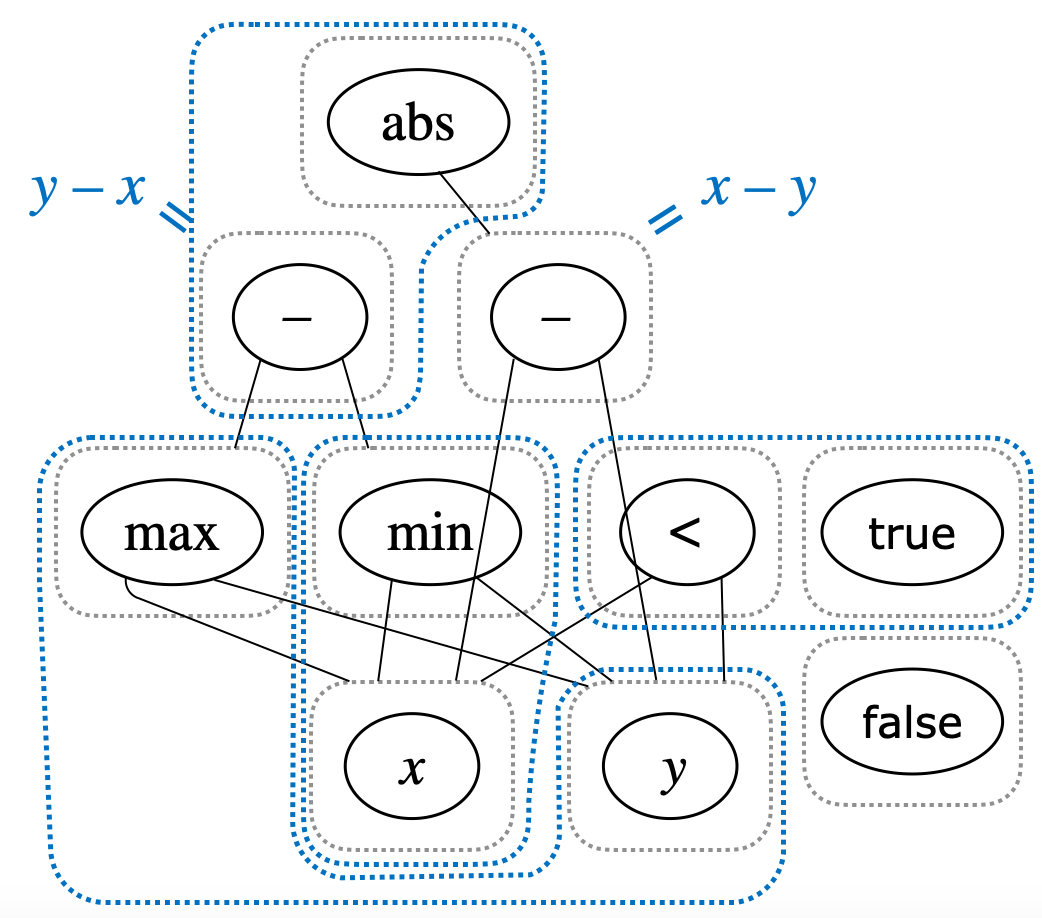} &
    \includegraphics[width=0.15\textwidth]{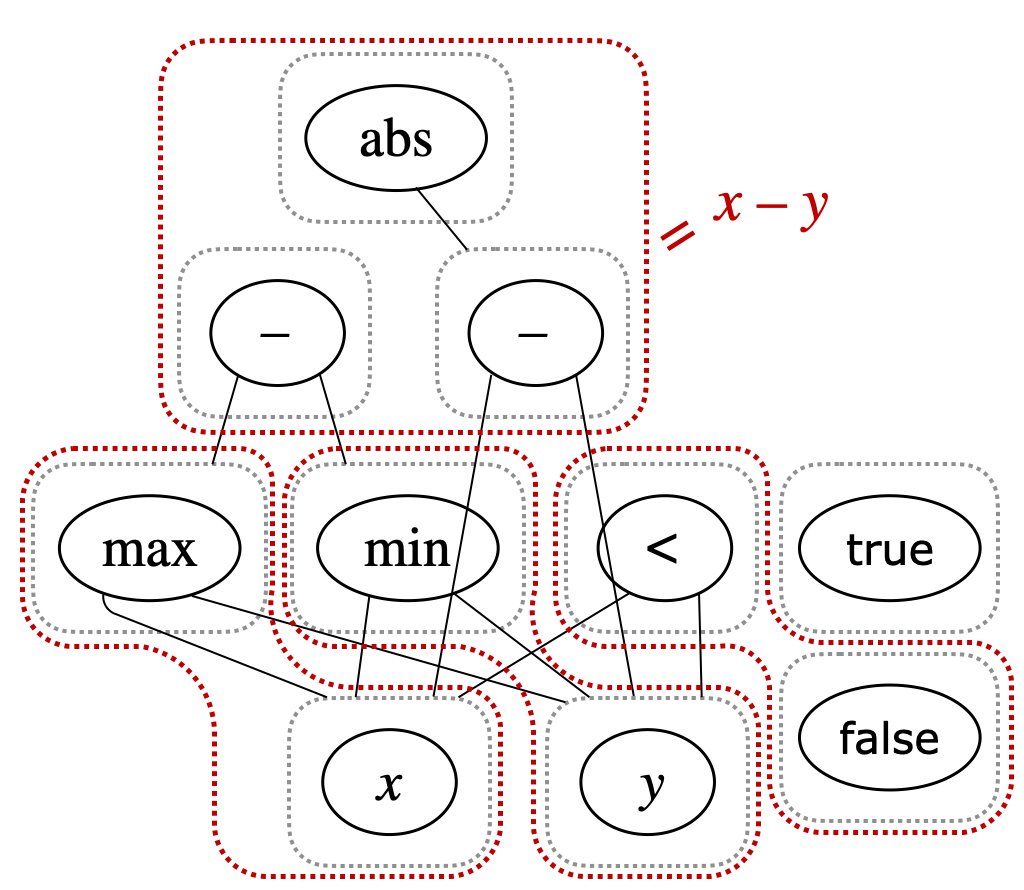}
    \\
    {\small (a)} & {\small (b)} & {\small (c)}
  \end{tabular}
  \caption{Proof of $\tsmallmax(x,y)-\tsmallmin(x,y)=|x-y|$.
  The e-nodes corresponding to the two terms are in the same e-class both in the blue layer (b) and in the red (c).
   It is important to note that the layers are overlain, and that the black nodes are shared; they are separated here for ease of perception.}
  \label{overview:egraph-max-min}
\end{figure}

This leads us to our proposed solution---%
\emph{Colored E-graphs}, whose
prime directive is to avoid duplication via sharing of the common terms, thus storing them only once when possible.
The e-graph structure becomes \emph{layered}:
the lowermost layer represents a congruence relation over terms that is true in all cases (represented, normally, as e-classes containing e-nodes).
On top of it are layered additional congruence relations that arise from various assumptions.
Going back to our example, the lowermost layer is
shown in \autoref{overview:egraph-max}(a),
containing the terms $\tmax(x, y)$, $x < y$, $\ttrue$ and $\tfalse$.
Layers corresponding to assumptions $x < y$ and $x \geq y$ are shown alongside it in \ref{overview:egraph-max}(b) and \ref{overview:egraph-max}(c).
To evoke intuition, we associate with each layer a unique \emph{color}, and paint their e-classes (dotted outlines, in depicted e-graphs) accordingly.
Conventionally, the lowermost layer is associated with the color black.
In the sequel we will use \cblue for $x<y$ and \cred for $x\geq y$ when referring to the example.
In the \cblue layer, $(x < y) \congblue \ttrue$ and
$\tmax(x, y) \congblue y$;
in the \cred layer, $(x < y) \congred \tfalse$
and $\tmax(x, y) \congred x$.
This is shown via the corresponding 
\cblue and \cred dotted borders.
\ref{overview:egraph-max}(d) shows a depiction where both colors are overlain on the same graph, which is a more faithful representation of the concept of colored e-graphs,
although this visualization is clearly not scalable to larger graphs.
In \autoref{overview:egraph-max-min} a larger graph can be seen that includes the terms $\tmax(x,y)-\tmin(x,y)$ and $|x-y|$, after a few more rewriting steps.
An overlain graph will be quite incomprehensible in this case, so the layers are shown separately; it can be easily discerned that $\tmax(x,y)-\tmin(x,y) \congblue |x-y|$
as well as $\tmax(x,y)-\tmin(x,y) \congred |x-y|$.

Observe that both additional layers, \cblue and \cred, utilize only the existing (black) e-nodes. 
Each additional color can be represented by simply specifying further unions of e-classes on top of those in the black congruence relation.
We say that each color's congruence $\cong_c$ is a \emph{coarsening} of the black congruence, $\cong$,
in the simple logical sense, ${\cong}\subseteq{\cong_c}$.
A crucial challenge to address is the implementation of operations (insert, union, congruence closure, e-matching)
efficiently while preserving this invariant as well as the standard e-graph invariants.
The colored layers require special support, as different e-classes may be united in the colored (non-black) layer but not in others (including the black relation). 
Notably, the black congruence relation can be implemented as a standard e-graph since all the necessary data structures are available to it.

Before diving into the design of colored e-graphs, it is better to start with their expected semantics.
One way to understand the semantics of colored e-graphs is by analogy to a set of clones, i.e. separate e-graphs $\mathcal{E}$;
One e-graph represents the base congruence $\cong$,
and one per color $c$ represents $\cong_c$.
All e-graphs in $\mathcal{E}$ conceptually represent the same terms partitioned differently into e-classes.
Thus, they have the same e-nodes, except that the choice of e-class id (the representative) may be different according to the composition of the e-classes.
We will call the e-classes of the color congruences \emph{colored e-classes}.
A union in the black layer (i.e. the original e-graph) is analogous to a union in \emph{all} of the e-graphs of the corresponding e-classes;
this maintains the invariant that every colored e-class is a union of (one or more) black e-classes.
A colored union, that is, a union performed in a colored layer, is just one union applied to the respective e-graph.
The colored e-graph semantics of the other operations---insertion, congruence closure, and e-matching---are the same as if they were performed across all clones.

A naturally occurring situation in equality saturation and exploratory reasoning tasks is that the e-graph is extensive, and each assumption induces a relatively small number of additional congruences.
Colored e-graphs are adapted to this scenario.
Each color layer corresponds to a narrow assumption, such as $x < y$ in our overview example, and will therefore have some additional unions, but not drastically many.
In such settings, the space reduction obtained by de-duplication of the black e-nodes, outweighs the overheads associated with bookkeeping of the colored e-classes.
With careful tweaks and a few optimizations, we show that we can keep these overheads quite modest.

For presentation purposes, we will describe the design and implementation of colored e-graphs in two steps.
We start with a basic implementation that is not very efficient but is effective for understanding the concepts and data structures;
then, we indicate some pain points, and move on to
\autoref{optimizations} to describe optimization steps
that can alleviate them.

In the basic implementation, all e-nodes reside in the ``black'' layer, which is represented by a ``vanilla'' e-graph from e-graphs library egg, including the operations listed in \autoref{background}.
The colored congruences do not have designated e-graphs of their own, and instead, the operations of union, congruence closure, and e-matching have \emph{colored variants}, parameterized by an additional color $c$ that are semantically analogous to the same operations having been applied, in clone semantics, to the e-graph associated with color $c$ in $\mathcal{E}$.
(For the time being, insertion does not have a colored variant, since insertions create e-nodes and all e-nodes are shared.)

\myparagraph{Colored union-find.}
In an e-graph, the union-find data structure holds all the e-class ids that were ever inserted into the e-graph.
Instead of having this information replicated per layer, colored e-graphs save a master copy onto which black unions are applied,
and on top of that \emph{smaller} union-finds,
one per color, whose elements are only the representatives of black e-classes that have been merged in each color.

\myparagraph{Colored e-matching.}
The e-class map is only saved for the black layer.
This is sufficient, because an e-class in color $c$ is always going to be a union of black e-classes, and all that is required for e-matching is finding e-nodes with a particular root (operator) in the course of the top-down traversal.
So the union can be searched on demand by collecting all the ``$c$-color siblings'' of the e-class and searching them as well.

\myparagraph{Colored congruence closure.}
In egg, the e-graph will fix the congruence invariant by going through the parents, re-canonising them, and find unions needed to complete the congruence by finding duplicates.
Potentially there could also be duplicates in the hash-cons, so parents are also looked up there.
As opposed to going through all e-classes during each hash-cons rebuilding, having to only go through the parents of e-classes that changed can be significant.
But, in colored e-graphs, there is no hash-cons for each color, so it is created on the fly.
Creating a hash-cons is done by iterating through all e-classes and collecting all e-nodes, and is therefore expected to be much slower.

\medskip
When using the above operations in the context of equality saturation, e-matching is applied for all colors
to discover matches for the left-hand sides of rules.
For each match, the right-hand side of the rule needs to be inserted into the e-graph and union-ed or color-union-ed with the left-hand side.
Inserting the e-nodes to the e-graphs makes them available to all layers.
We assume that the mere \emph{existence} of a term in an e-graph does not in itself have the semantics of a judgement---it is only the placing e-nodes in the same e-class that asserts an equality.
Thus, under this assumption, inserting all the e-nodes to the e-graph is sound.
However, in the presence of many colors, and thus many colored matches, the result would be a large volume of e-nodes that are in black e-classes of size 1, because they
were created just to serve one of the colors.
As opposed to a standard, single e-graph where merging e-classes shrinks the space of e-nodes (because non-equal e-nodes may become equal as a result of canonization),
in colored unions it is required that the e-graph maintain both original e-classes, thus losing this advantage.
This can put a growing pressure on subsequent e-matching and rebuild operations \emph{in all colors}.

\section{Optimizations}
\label{optimizations}

In this section, we explain in more detail some important optimizations that address various challenges encountered when applying operations such as e-matching and rebuilding in colored e-graphs. 
These optimizations aim to mitigate the negative impact of having e-nodes from multiple colored layers, inefficient rebuild processes, duplicate results in e-matching, and the re-adding of e-nodes.
We will first explain these problems in more detail, and then offer some solutions

Rebuilding in colored e-graph, as discussed in \autoref{overview}, can be significantly slower compared to a separate, minimized e-graph.
The two main causes are that building a colored hash-cons requires going over all the e-classes, and that the colored e-graph contains duplicate e-nodes compared to a separate minimized one.

E-matching for each color may produce duplicate results due to the e-graph not being minimized according to the color's congruence relation;
that is, colored-congruent terms are not always merged under a single e-class.
To illustrate this, consider a simple e-graph
representing the terms $1\cdot 1$, $1\cdot x$, $1\cdot y$, and $x\cdot y$.
Introduce a color, \cblue, where $x \congblue y$.
A simple pattern such as $1\cdot ?v$ would have three matches, with assignments $?v \mapsto 1, ?v\mapsto x, ?v\mapsto y$.
If the \cblue layer were a separate e-graph, $x$ and $y$ would have been in the same e-class,
so one of the matches here is redundant (as far as the blue layer is concerned).
Of course, in the black layer they are different matches; the point is, that many terms are added to the graph only as a result of a colored match,
so matching them in the black e-graph is mostly useless to the reasoner.
On the other hand, their \emph{presence} in the black layer means they cannot ever be union-ed, leading to duplicate matches, as seen above, even in the respective colored layer(s).

When inserting e-nodes to the e-graph, the hash-cons is used to prevent duplication based on it being canonized.
Adding an e-node from a colored conclusion, that is the e-matching relied on the colored congruence relation, does not benefit from canonization.
In fact, each e-node $f(x_1,\dots,x_n)$ has a multitude of equal black representatives that are $\congblue$-equivalent. 
Each child $x_i$ in the e-node can be presented by any black id such that $e \in [x_i]_b$, so there are $\prod_i |[x_i]_b|$ representations.
And, each representative may be re-added to the e-graph without \cblue canonization.

To address these issues, we present a series of optimizations to the colored e-graph data-structure and the procedures. 
These optimizations aim to reuse the ''root`` layer as much as possible, in both memory use and also in compute.
Thus, we can achieve a memory efficient, but effective colored e-graph. 

\subsection{Data-structure optimizations}
\myparagraph{Colored e-nodes.}
Adding e-nodes resulting from colored e-matches to the root e-graph is the source of many inefficiencies in the simple implementation as described in \autoref{overview}.
Instead, the optimized version has \emph{colored e-nodes} as well; that is, e-nodes that are introduced following a colored match are associated with that color.
Each colored layer includes additional specialized colored hash-cons and colored e-class map that contains colored e-nodes and colored parents. 
These structures should only hold the difference from the ``black'' structures, 
reusing as much as possible from the ``root'' layer.

Using the new colored e-classes and hash-cons, colored e-matching and the rebuild process are done once per color, and only need to go through the ``black'' e-classes with the addition of the colored e-classes of the current color.
An important quality of the colored hash-cons is that it is canonized to its specific color.
Whenever a new colored e-node that added, it is checked against this canonical colored hash-cons to prevent duplicate insertion.
We still might add unnecessary colored e-nodes, as the ``root'' layer might contain an equivalent e-node, but we will at most add each colored e-node one unnecessary time. 


\myparagraph{Pruning.}
Recall that having only coarsening relations means that any result found in a finer (black) relation is also true for the coarsened relation(s).
And so, following black unions, some of the colored e-nodes could actually become subsumed by the ``black'' e-nodes during the equality saturation run.
To overcome these redundant e-nodes, we present an efficient deferred pruning method.

Normal e-graph minimization relies on having all e-nodes canonized.
A colored e-graph does canonize all e-nodes (including the black ones) during rebuild, by constructing the colored hash-cons.
Thus, the redundant e-nodes in the colored e-class map will not be present in the colored hash-cons, and can therefore be removed.
While pruning is promising, its usability is limited to cases where the colored e-node will not be added back.

\myparagraph{Colored minimization.}
Another improvement is having multiple colored e-nodes (of the same color) in a single (black) e-class. 
Previously all the additional colored e-nodes had to be in their own e-classes, as no black unions would be performed on them.
In the optimized version, two black e-classes can be merged if both contain colored e-nodes of the same color and are in the same colored e-class (of the same color).
Thus, an invariant is kept that each colored equality class has at most one black e-class containing colored e-nodes.

\subsection{Procedures optimizations}
\myparagraph{Rebuild.}
When rebuilding, we first reconstruct the congruence relation of the ``root'' layer.
Even though a color, for example {\color{blue}blue}, will need to rebuild its own congruence, it still holds that ${\cong}\subseteq{\congblue}$.
So, any union induced by ${\cong}$ can be applied to the {\color{blue}blue} relation.
To understand the implications, consider the e-graph representing the terms $x$, $y$, $f(x)$, $f(y)$, $f(f(x))$, and $f(g(y))$ where the {\color{blue}blue} color contains the additional assumption that $g(y) \congblue f(y)$.
If we union $x$ and $y$, the black congruence will include $f(x) \cong f(y)$  which also holds in the blue relation.
But, the depth of the blue rebuilding is once more, as $g(y) \congblue f(y)$, we need to conclude the $f(f(x)) \congblue f(g(y))$.
This demonstrates how reusing the black relation is useful; the colored hash-cons is rebuilt once even though the rebuilding depth is two.

\myparagraph{E-match.}
In e-matching a similar optimization is applied.
Any conclusion of e-matching on the root layer is also applicable to any higher layer in the e-graph.
Thus, during e-matching, a pattern matched against the current layer should not be repeated for a higher layer.
The e-matching is then implemented starting with a standard top-down traversal for the root layer, but with the additional on demand colored e-matching.
That is, when traversing an e-class that has colored siblings (\ie is contained in a larger, colored e-class) that can be matched with the current pattern being searched,
e-matching will be \emph{forked}, ``jumping over'' to that sibling,
and continuing to search downward.
This way, all forked matching paths are unique, as at least one (different) e-class is chosen at each fork.
Although all expected matches are found, and there are no duplicate paths, duplicate colored matches will still be present due to the e-graph not being fully minimized.

\section{Evaluation}
\label{eval}




In this section, we evaluate the performance and effectiveness of our colored e-graphs and the different optimizations we propose.
For this purpose we implemented two versions of colored e-graphs containing different improvements described in \autoref{optimizations}.
The simple version only uses procedural improvements, while the optimized version uses all optimizations.
A small caveat in our colored e-graph implementation is that layered union-finds are not implemented directly.
Instead, it is simulated by creating a full copy of the original union-find for each new color.


\subsection{Objectives and Evaluation Method}
The objective of our evaluation is to assess the potential usefulness of colored e-graphs for exploratory tasks, specifically focusing on their ability to efficiently run equality saturation for multiple assumption sets simultaneously. 
To accomplish this we designed an experiment consisting of multiple test cases, each containing a distinct set of rewrite rules and proof goals. 
Our evaluation metrics include measuring the size of the e-graph created, and the time taken for equality saturation.
To the best of our knowledge, a purely e-graph based automated theorem prover does not exist, and theory exploration tools have limited support for conditions.
Therefore, for the experiments, we constructed an equality saturation-based prover (based on the code from~\cite{thesy}) that incorporates an automatic case-splitting mechanism. 

The case-splitting mechanism is only used when it will potentially contribute to progress in equality saturation---that is, when it enables additional rewrite rules that were previously blocked.
Each case split creates additional, coarsened e-graphs.
As a baseline, this prover uses a set of e-graphs (clones) to represent the different congruence relations.
To more closely simulate an exploratory reasoning scenario, we consider the accumulated size of all the clones in our comparison.

We experiment on inductive proof test suites introduced in \cite{VMCAI2015:Reynolds}, which are also used in \cite{thesy}.
Since the instances are relatively small, we introduced a slight variation:
typically, only the terms occurring in the proof goal are used as input to the equality saturation process.
In our experiments, for each proof goal, we identified all other benchmarks within the same test suite that have similar goals; that is, have a common vocabulary and common rewrite rules. 
The prover thus explores the e-graph of all the proof terms emerging from these starting set, which gives rise to larger e-graphs and more significant measurements.
Importantly, the prover does not stop even if it proves the goal early, but continues to explore the space until saturation or resource cap.

All the experiments were conducted on 64 core AMD EPYC 7742 processor with 512 GB RAM, ensuring consistent testing conditions for accurate comparisons and measurements.

\subsection{Experimental Setup}

Using the modified prover, each test case is run, e-graph size, and run-times.
The measurement of e-graph sizes involved counting the number of e-nodes present in the e-graph.
For the colored layers, we specifically measured the additional colored e-nodes in the e-graph for each layer. 
In contrast, for the separate set of e-graphs, we counted the e-nodes in the original e-graph as well as each of the coarsened e-graphs.
To manage memory usage during the evaluation, we incorporate the Cap library, a Rust library specifically designed to limit memory usage. 
Each test case was limited to 32 GB of memory and 1 hour run-time.

In our evaluation, we ran experiments using 
a basic implementation of colored e-graphs as described in \autoref{overview}, but without the data-structure optimizations (``monochrome e-nodes''), and the fully optimized one.
We compare each of them against a baseline of using separate e-graphs.

As for other forms of ablation, in our experiments, we also measured the optimized implementation without pruning.
It shows results that are very similar to the version that includes pruning.
This is expected because in this experimental setup most of the colored e-nodes will be re-added by running the rewrite rules again.

By conducting experiments with these three different flavors of e-graphs with additional assumptions, we aim to compare their performance and measure their impact on e-graph size and run-time. 
The results of these experiments provide 
insights into the advantages and limitations of each approach.

\subsection{Results}

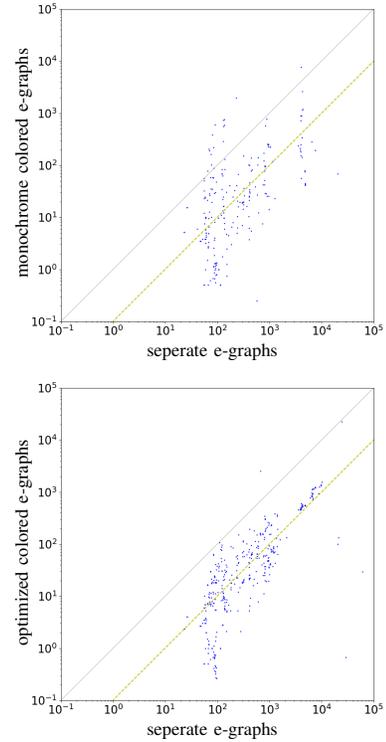
\begin{figure}
  \centering
  \newcommand\axislab[1]{\fontsize{7pt}{9pt}\selectfont{#1}}
\begin{tikzpicture}
  \node(mono)[inner sep=0,draw] {
   \includegraphics[width=0.25\textwidth,
                    trim=20mm 20mm 20mm 20mm] 
     {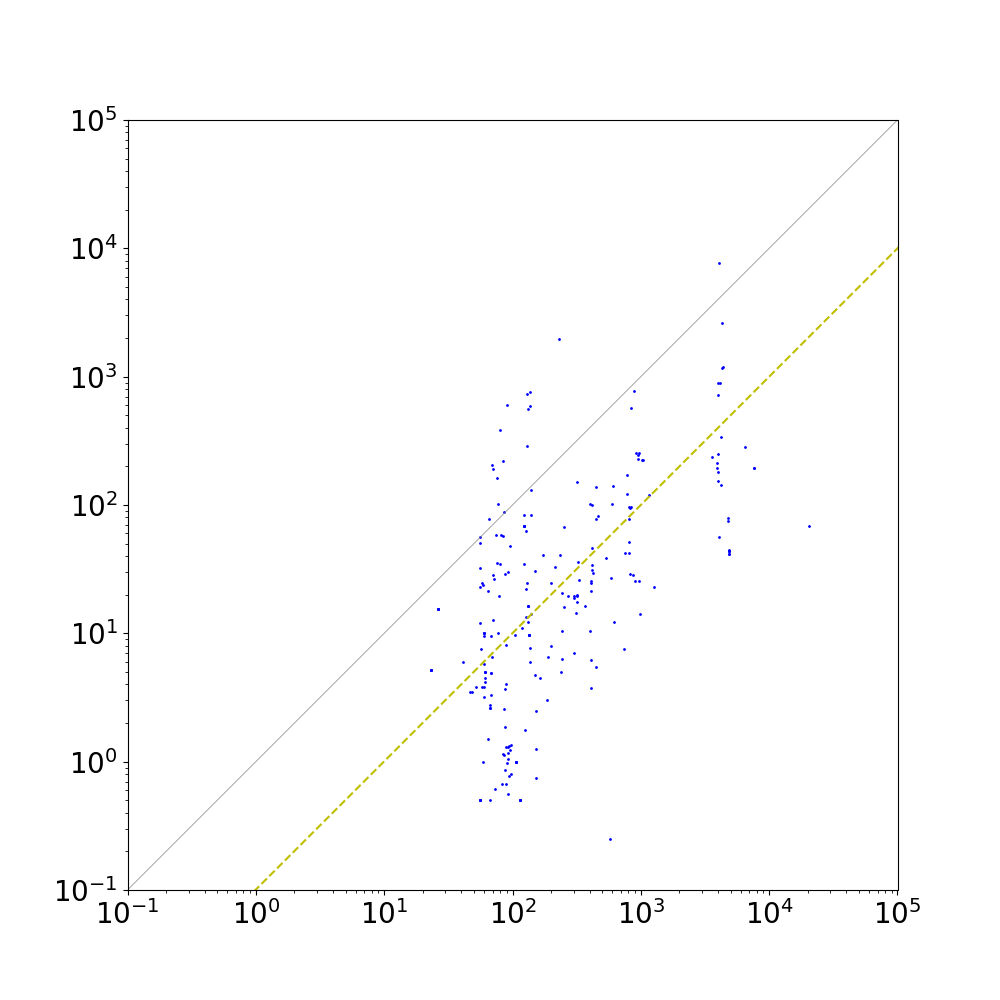}
     };
  \node(opt)[below=5mm of mono, inner sep=0] {
   \includegraphics[width=0.25\textwidth,
                    trim=20mm 20mm 20mm 20mm] 
     {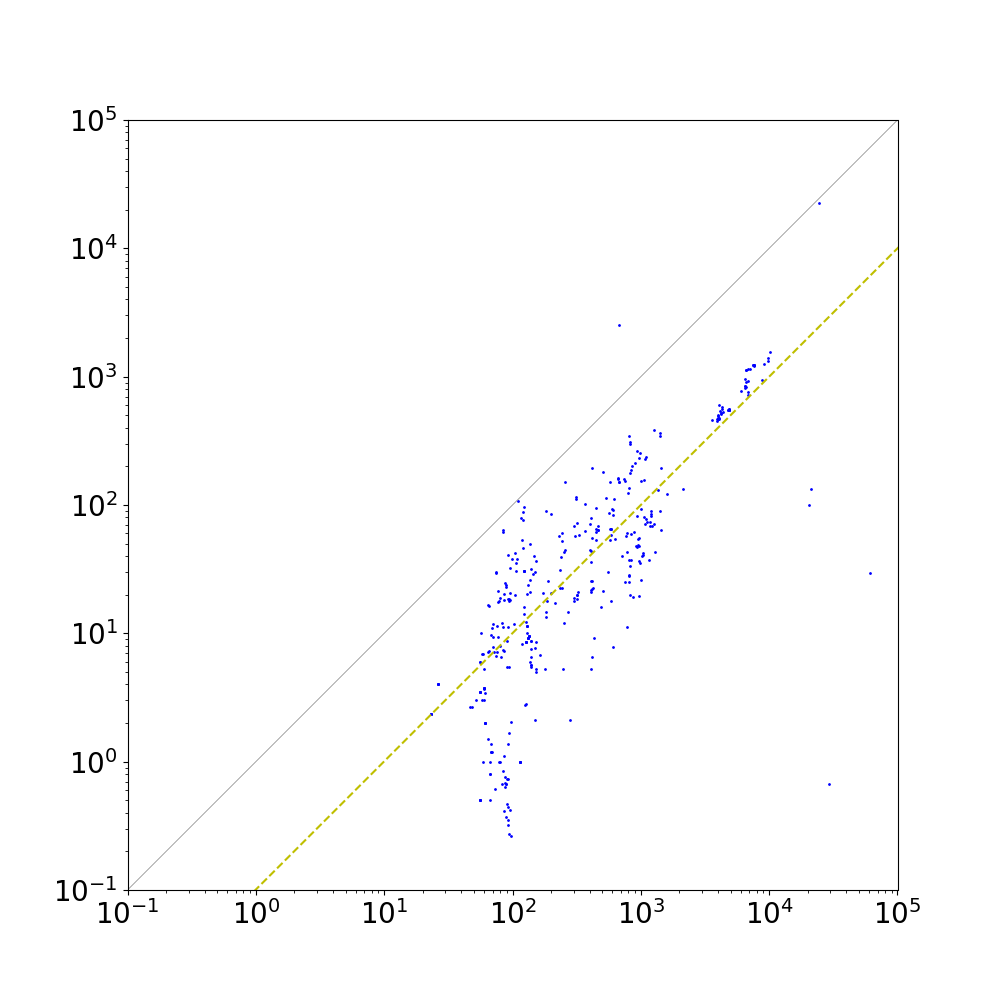}
     };
  \node[below=0mm of mono] {\axislab{seperate e-graphs}};
  \node[left=0mm of mono, rotate=90, anchor=south] {\axislab{monochrome colored e-graphs}};
  \node[below=0mm of opt] {\axislab{seperate e-graphs}};
  \node[left=0mm of opt, rotate=90, anchor=south] {\axislab{optimized colored e-graphs}};
\end{tikzpicture}
  \caption{Size comparison: relative e-node overhead in clones \vs color e-graph variants.}
  \label{fig:normalizedsize}
\end{figure}

In our setup, all assumptions emerge from case splits done by the prover.
We filter out cases were no case splits were applied, since these have no assumptions and thus colored e-graphs have no impact.

For each benchmark instance, we measure the \emph{relative e-node overhead} as the number of additional e-nodes that are required, normalized by the number of different assumptions.
That is, $(|\textrm{total e-nodes}| - |\textrm{base e-nodes}|) / |\textrm{assumptions}|$.
``Base e-nodes'' represent the contents of the graph before case splits.
(For the colored e-graph with monochrome e-nodes we use the base e-nodes present in the separate e-graphs case.)
\autoref{fig:normalizedsize} summarizes the results, pitting colored e-graphs (with and without colored e-nodes) against the baseline of separate clones.
In some cases one configuration times out or runs out of memory, while the other does not;
we only compare cases where both configurations finished the run successfully.
In both comparisons, we see roughly around 10$\times$ lower overheads, where in the monochromatic case samples are more dispersed around the y axis, and the optimized case shows clear advantage to the colored e-graph implementation.

\begin{figure}
  \centering
  \newcommand\axislab[1]{\fontsize{7pt}{9pt}\selectfont{#1}}
\begin{tikzpicture}
  \node(mono)[inner sep=0] {
   \includegraphics[width=0.3\textwidth] 
     {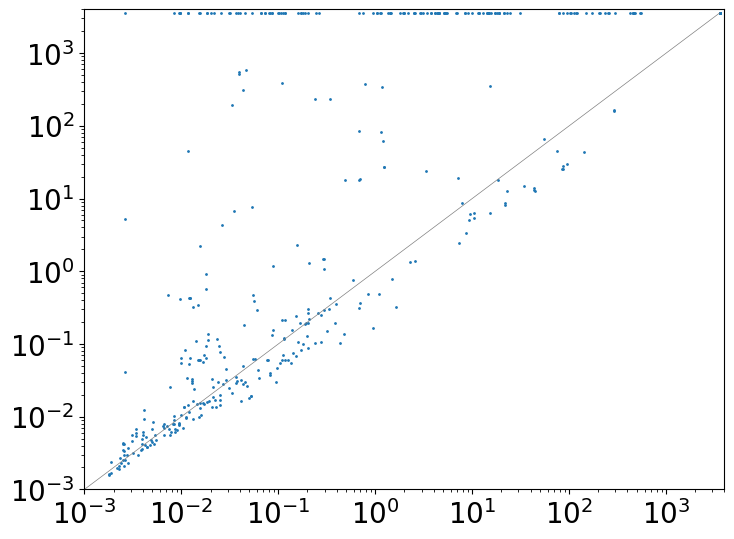}
     };
  \node(opt)[below=10mm of mono, inner sep=0] {
   \includegraphics[width=0.3\textwidth] 
     {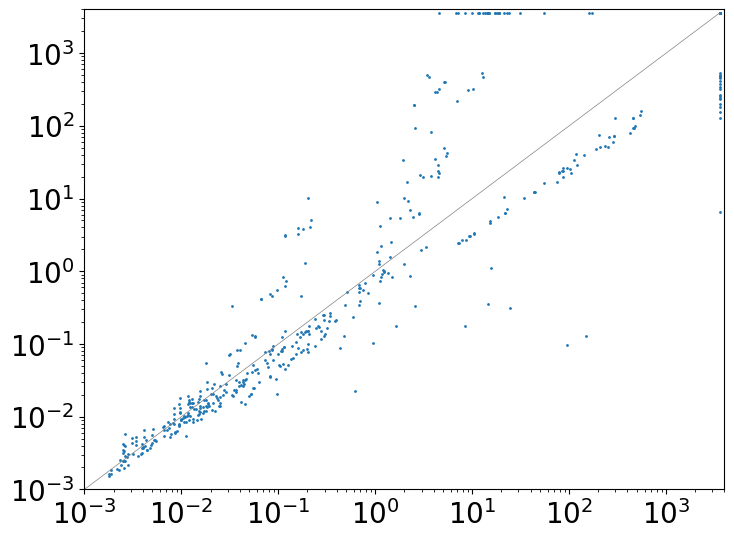}
     };
  \node[below=0mm of mono] {\axislab{seperate e-graphs}};
  \node[left=0mm of mono, rotate=90, anchor=south] {\axislab{monochrome colored e-graphs}};
  \node[below=0mm of opt] {\axislab{seperate e-graphs}};
  \node[left=0mm of opt, rotate=90, anchor=south] {\axislab{optimized colored e-graphs}};
\end{tikzpicture}
    \caption{Run-time comparison: run-time of clones vs. color e-graphs}
    \label{fig:runtime}
\end{figure}
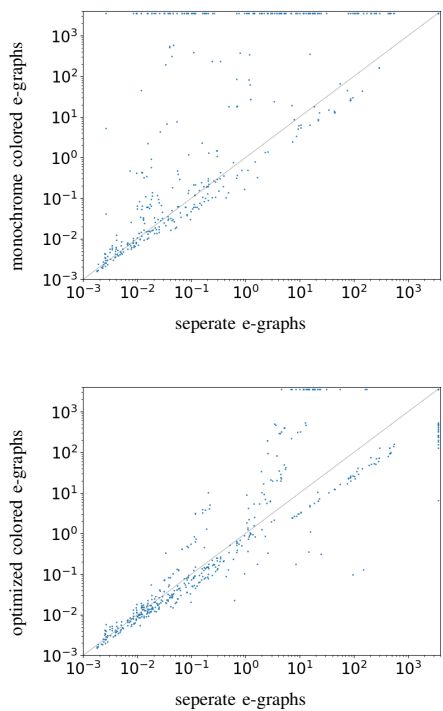

Run-time is measured as the the total run-time for completed test cases, and 1 hour for cases that timed out.
We do not include runs that did not finish due to out-of-memory exceptions (we report the latter separately).
As can be seen in \autoref{fig:runtime}, the monochrome e-nodes lead to many timeouts,
whereas the optimized case exhibits running times similar to separate clones.
This is in line with our expectation:
colors provide lower memory sizes at the expense of run-time.

Finally, in \autoref{tab:runtimesuite} we present the amount out-of-memory exceptions, amount of timeout exceptions, and total run-time for each configurations and test-suite.
The monochrome e-graph, as expected, has many timeout and out-of-memory exceptions. 
Even though it has more errors then the other e-graph versions, it still has much longer run-times.

The optimized e-graph has results much more similar to those of the separate e-graphs, both in run-time and amount of exceptions.
The main differences can be seen in errors moving from out-of-memory to timeouts.
One special case is in the \funcname{leon-heap} test suite we see a difference of 12 timeout exceptions in the optimized version, where all the 12 timeouts appear in split depth 4.
This difference rises from small implementation differences between the colored e-graph and separate e-graphs which can lead to a different amount of assumptions found during TheSy's prover run.
Usually both separate e-graphs and colored e-graphs have a similar number of assumptions, but there are a few outliers, one of which in the \funcname{leon-heap} test suite.
Looking deeper, this case runs out of memory at a split depth of 4; at depth 3, we see this disparity, with a total of 1426 assumptions for separate e-graphs but 2676 assumptions for colored e-graphs.

\begin{table}[t]
    \centering
    \caption{Run-time and exceptions for each test suite}
    \label{tab:runtimesuite}
    \resizebox{0.5\textwidth}{!}{
    \begin{tabular}{llrrr}
    \toprule
     &  & Run-time & OOM & Timeout \\
    Type & Test Suite & {\fontsize{7pt}{6pt}\selectfont (seconds)}  &  &  \\
    \midrule
    \multirow[t]{6}{*}{Seperate} & clam & 33.1 & 0 & 0 \\
     & hipspec-rev-equiv & 2.6 & 0 & 0 \\
     & hipspec-rotate & 7437.7 & 2 & 0 \\
     & isaplanner & 16113.7 & 2 & 58 \\
     & leon-amortize-queue & 187209.7 & 52 & 2 \\
     & leon-heap & 216.9 & 0 & 0 \\
    \cline{1-5}
    \multirow[t]{6}{*}{Monochrome} & clam & 393.5& 0 & 3 \\
     & hipspec-rev-equiv & 43200.7 & 12 & 12 \\
     & hipspec-rotate & 1970.5 & 0 & 13 \\
     & isaplanner & 73739.2 & 20 & 136 \\
     & leon-amortize-queue & 39600.3 & 11 & 48 \\
     & leon-heap & 12006.0 & 3 & 29 \\
    \cline{1-5}
    \multirow[t]{6}{*}{Optimized} & clam & 4.7 & 0 & 0 \\
     & hipspec-rev-equiv & 42.0 & 0 & 0 \\
     & hipspec-rotate & 78.2 & 0 & 2 \\
     & isaplanner & 19537.0 & 2 & 57 \\
     & leon-amortize-queue & 10843.8 & 3 & 50 \\
     & leon-heap & 791.4 & 0 & 12 \\
    \bottomrule
    \end{tabular}
}
\end{table}

\section{Related Work}
\label{related}

\myparagraph{Theory exploration and its applications.}
Interest in exploratory reasoning in the context of functional calculi started with IsaCoSy~\cite{JAR2010:Johansson}, a system for lemma discovery based in part on CEGIS~\cite{ASPLOS2006/Solar-Lezama}.
In a seminal paper, QuickSpec~\cite{JFP2017:Smallbone}
propelled applicability of such reasoning for inferring specifications from implementations
based on random testing,
with deductive reasoning to verify generated conjectures~\cite{ICAD2013:Claessen,ITP2017:Johansson}.
TheSy~\cite{thesy} and Ruler~\cite{DBLP:journals/pacmpl/NandiWZWSASGT21} have both incorporated e-graphs to some extent in the exploration process: they are used to speed up equivalence reduction of the space of generated terms, and, in~\cite{thesy}, also the filtering and qualification phases using symbolic examples.
The evaluation of the latter shows quite clearly that case splitting is a major obstacle to symbolic exploratory reasoning,
due to the large number of different cases and derived assumptions.

In the area of conditional rewrite discovery,
Speculate~\cite{DBLP:conf/haskell/BraquehaisR17}
naturally builds on the techniques from QuickSpec and depends on property-based testing techniques to generate inputs that satisfy some conditions.
SWAPPER~\cite{DBLP:conf/fmcad/0002S16} is a relatively early example of exploring using SyGuS with a data-driven inductive-synthesis approach with emphasis on finding rules that are most efficient for different problem domains.
It requires a large corpus of similar SMT problems to operate.


\myparagraph{Other e-graph extensions.}
E-graphs were originally brought into use for automated theorem proving~\cite{JACM2005:Detlefs},
and were later popularized as a mechanism for implementing low-level compiler
optimizations~\cite{POPL2009:Tate},
by extending them with ``$\varphi$-nodes'' to express loops.
Relational e-matching~\cite{DBLP:journals/pacmpl/ZhangWWT22}
makes use of Datalog semina\"ive evaluation
to harness the power of query planning in database systems.
Subsequently, Datalog-powered e-matching has been recently fused with core Datalog semantics to allow richer logic programming by exposing equality saturation as a building block in a framework called egglog~\cite{DBLP:pldi23/Zhang}.
Since Datalog is based on Horn clauses, this meshes very well with conditional rewriting.
It should be noted, though, that it is still a monotone framework, and does not allow backtracking or simultaneous exploration of alternative assumptions.

ECTAs~\cite{Koppel22,Koppel23} are another, related compact data structure that extends e-graphs, Version-Space Algebras~\cite{DBLP:conf/icml/LauDW00,DBLP:journals/ml/LauWDW03}, and Finite Tree Automata~\cite{DBLP:journals/pacmpl/0001M17},
with the concept of ``entanglement'';
that is, some choices of terms from e-classes
may depend on choices done in other e-classes.
Since the backbone of ECTAs is quite similar to an e-graph, the colors extension is applicable to this domain as well.

\myparagraph{Uses of e-graphs in SMT.}
E-graphs are a core component for equality reasoning in SMT solvers~\cite{TACAS2008:DeMoura,DBLP:conf/tacas/BarbosaBBKLMMMN22}, in most theory solvers such as QF\_UF, linear algebra, and bit-vectors.
E-matching is also used for quantifier instantiation~\cite{DBLP:conf/tacas/NiemetzPRBT21}, which is, in its essence, an exploratory task and requires efficient methods~\cite{DBLP:conf/cade/MouraB07}.
In these contexts, implications and other Boolean structures are treated by the SAT core (in CDCL(T)), and the theory solver only handles conjunctions of literals.

\section{Conclusions}
\label{conclusions}


In conclusion, this paper has introduced the concept of colored e-graphs as a memory-efficient method for maintaining multiple congruence relations in a single e-graph. 
It provides support for equality saturation with additional assumptions over e-graphs, thereby enabling efficient exploratory reasoning of multiple assumptions simultaneously.
The development of several optimizations based on the egg library and deferred rebuilding, and subsequent evaluation has validated our approach, demonstrating a significant improvement in memory utilization while ensuring competitive run-time performance compared to the baseline.

This work thus serves as a stepping stone, advancing the current state of the art and setting a foundation for works on exploratory reasoning tools and techniques. 
By extending e-graph capabilities, we hope to drive new innovation in the realm of symbolic reasoning and its applications.

\bibliographystyle{IEEEtran}
\bibliography{bib}

\end{document}